\documentclass[a4paper,fleqn]{cas-sc}
\usepackage[sort&compress,square]{natbib}

\usepackage{import}
\usepackage{mathtools}
\usepackage{subfig}
\usepackage{amssymb}
\usepackage{stackengine}
\usepackage{scalerel}
\usepackage{xcolor}
\usepackage{graphicx}
\usepackage{lineno}
\DeclarePairedDelimiter\abs{\lvert}{\rvert}%
\DeclarePairedDelimiter\norm{\lVert}{\rVert}%

\newcommand{\com}[1]{\textcolor{black}{#1}}
\newcommand\openbigstar[1][0.7]{%
  \scalerel*{%
    \stackinset{c}{-.125pt}{c}{}{\scalebox{#1}{\color{white}{$\bigstar$}}}{%
      $\bigstar$}%
  }{\bigstar}
}

\makeatletter
\let\oldabs\abs
\def\abs{\@ifstar{\oldabs}{\oldabs*}}
\let\oldnorm\norm
\def\norm{\@ifstar{\oldnorm}{\oldnorm*}}
\makeatother

\begin{document}

\let\WriteBookmarks\relax
\def\floatpagepagefraction{1}
\def\textpagefraction{.001}

\title[mode=title]{Can confined mechanical metamaterials replace adhesives?}
\shortauthors{Athanasiadis A., Dias M., Budzik M.}
\shorttitle{Confined mechanical metamaterials}
\author[1]{Adrianos E. F. Athanasiadis}
\author[2]{Marcelo A. Dias}
\cormark[1]
\ead{marcelo.dias@ed.ac.uk}
\author[1]{Michal K. Budzik}
\cormark[1]
\ead{mibu@mpe.au.dk}
\address[1]{Lab. Mech. Phys. Solids (LAMPS), Department of Mechanical and Production Engineering, Aarhus University, Inge Lehmanns Gade 10, 8000 Aarhus C, Denmark}
\address[2]{Institute for Infrastructure \& Environment, School of Engineering, The University of Edinburgh, EH9 3FG Scotland, UK}
\cortext[cor1]{Corresponding authors}
\maketitle
\linenumbers
\begin{abstract}
The subject of mechanical metamaterials has been gaining significant attention, however, their widespread application is still halted. Such materials are usually considered as stand-alone, vis-\`a-vis all characteristic length scales being associated solely with geometry of material itself. In this work we propose novel application of mechanical metamaterials as interface regions joining two materials with potential of replacing bulk adhesives. This idea leads into paradigm shifts for both metamaterials and adhesive joints. In specific, we outline methodology for testing and evaluating confined lattice materials within fracture mechanics framework. The theoretical and numerical approaches are inter-winded, revealing a set of critical parameters that needs to be considered during design process. Lattices that are stretching and bending dominated are explored and failure maps are proposed, indicating susceptibility to a certain failure mode depending on level of confinement and characteristic dimension of each lattice's unit cells. 
\end{abstract}





\section{Introduction}

In fast-growing industrial sectors for infrastructure engineering, such as transportation and energy, drives for efficiency and functionality inevitably rely on a deep understanding of confined materials and adhesive joints~\citep{zuo2020review,xu2020effects}---\emph{i.e}, joining or sealing two similar or dissimilar materials is paramount to progress. Examples are found in glass fibre composite (GFRP) parts of a wind turbine rotor blades~\citep{DNVGL_ST_0376}, trending glass bridges connected to metallic or wooden frames~\citep{vallee2016design}, and timber construction with in-rod mounting~\citep{barbalic2021short}. In many of these instances, the assumption of idealised and seamless joints are commonly far from ideal to be considered in calculations, and disregarding their intrinsic characteristic dimensions and constitutive behaviour is more often than not leading fundamental design flaws. Complex interactions amongst different joint constituents, mediated by finite thickness adhesive layer, does play a fundamental role in structural integrity, \emph{i.e.}, the behaviour during loading and failure~\citep{pardoen2005constraint,van2019tensile}.

\begin{figure}
	\centering
	\includegraphics[width=\textwidth]{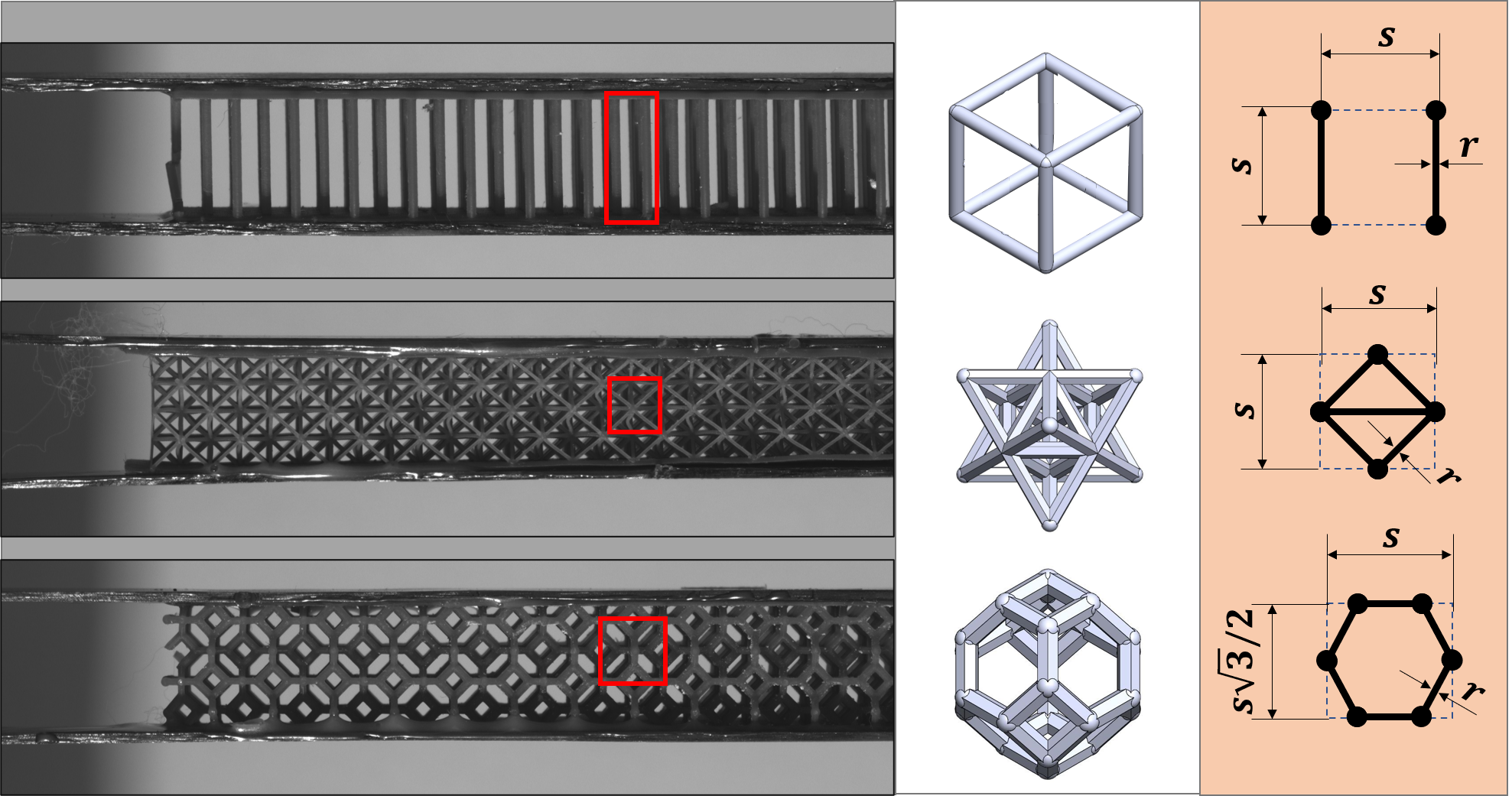}
	\caption{\com{For illustration purpose,} proof-of-concept of manufacturability of the specimens with interface unit cells studied in this work. The specimen consisted of carbon fibre/epoxy matrix (CFRP) composites joined by: (top row) pillar, (middle row) octet truss, and (bottom row) Kelvin cell interfaces.}
	\label{fig:exp}
\end{figure}

It is now recognised, that the thicker and more compliant adhesive layers can offer a set of advantages that should no longer be ignored, \emph{e.g.}, enhanced fracture toughness and damage tolerance, or better damping and insulation properties \citep{banea2015effect,pascoe2020effect}. Moreover, in many applications, thick layers cannot be avoided and they help accommodating manufacturing tolerances~\citep{zuo2020review}. However, this comes at a price of volume-to-weight penalty, reduced stiffness and severe stress gradients close to the corners and edges created along the adhesive material surface~\citep{mohammed2000cohesive,akisanya2003initiation,LopesFernandes2021Stiff}. Exploration of confined lattice structures is correlated and consistent with related literature in sandwich structures. \com{In that regard, an important contribution has been made towards exploring the fabrication and static properties of simple lattice structures, against the more often used foams or honeycomb cell cores~\citep{wadley2003fabrication}.} Within this framework, the main role of the lattice core is to ensure distance between joined materials, thus increasing bending stiffness by using minimum of material~\citep{song2019topology}. \com{This is indeed a case study for the optimisation of sandwich structures~\citep{triantafillou1987minimum,ashby2000metal}.} The role of the bondline is in many aspects, if not all, very different than that of sandwich core materials. Bondlines are used and designed to ensure structural integrity and maximise efficiency of load transfer between joined materials. Rather than being constrained by the aforementioned drawbacks of homogeneous bondlines, and search for optimal bending stiffness-to-weight-ratio, characteristic for sandwich structures, the primary aim of this work is to introduce concept of an interface with high volume-to-weight ratio, where the mechanical performances are not sacrificed, and with properties that could be carefully tailored along any direction and critical localisation. With emerging additive manufacturing technologies, our focus turns into architected interfaces with the scope of merging mechanical matematerials into joint designs. \com{Although the scope of this work is exclusively theoretical, we depict in Fig.~\ref{fig:exp} three possible specimens of architected interfaces with the sole purpose to introduce the concept. In these examples, the high strength carbon fibre composite adherends were joined through the $10$ mm bondlines obtained by SLA additive manufacturing.}
  
\com{In this work, we focus on modelling the three types of structures shown in Fig.~\ref{fig:exp}:} (i) an array of pillars; (ii) an orthogonal projection of the octet truss, resulting into a triangular tessellation; finally, (iii) inspired by the Kelvin cell, we investigate the hexagonal tiling. \com{However, to make the calculations less demanding, but keeping the scope intact, we consider structures embedded in 2-dimensions.}
The simplest of them all, the pillar interface, has already received significant attention in regards to bonding and materials interface design~\citep{zhou2013controllable,heide2020mechanics,kim2020designing}. This configuration aims at providing a benchmark for higher degree of complexity microstructures inspired by three-dimensional unit cells, specifically the tetrahedral-octahedral (octet truss) and the bitruncated cubic honeycomb (Kelvin cell). The three geometries will be further explored in the following sections and further merged into a common framework with the joined materials. Results of theoretical developments will finally be compared with numerical model. \com{Specifically, we aim at building a consistent theoretical framework that unifies metamaterial interface with composite adherents that will result in failure load predictive capabilities.}

\section{Effective mechanical properties of architected interfaces}

We begin by analysing the rigidity of the selected geometries, as shown in the last column of Fig.~\ref{fig:exp}. These are identified through Maxwell's rule~\citep{maxwell} and its generalisation, as formulated by~\citep{pellegrino1986matrix}. Maxwell's rule in two-dimensions states that a frame is rigid when $B-2J+3\geq0$, where $B$ and $J$ stand for, respectively, the number of bars and joints of the frame in question. This can be further generalised as $B-2J+3=S-M$, where $S$ counts the states of self-stress and $M$ the number of mechanisms. For the case of pillars (see first row of Fig.~\ref{fig:exp}), we include one bar ($B=1$) and two joints ($J=2$) hence, the unit cell is isostatic ($B-2J+3=0$). Similarly, for the triangular lattice (2D analogy of the octet truss as shown in the second row of Fig.~\ref{fig:exp}), we include five bars and four joints hence, the unit cell is isostatic ($B-2J+3=0$). For the hexagonal lattice (depicted in the last row of Fig.~\ref{fig:exp}), we include six bars and six joints hence, the resulting frame is a mechanism ($B-2J+3<0$). Since pillars and triangular lattice are isostatic frames, their members are only loaded axially, and therefore are behaving as stretching dominated geometries. Moreover, when the joints of the mechanism formed by the hexagonal lattice are locked in place, the unit cell is loaded with bending moment, and therefore is behaving as bending dominated geometry~\citep{Zheng2014}. 
The design variable of a given lattice structure consisting of struts is taken to be the inverse of the slenderness parameter, which is given by the aspect ratio $\tilde{\ell}=r/s$, where there radius of the strut $r$ and the unit cell length $s$ are illustrated in Fig.~\ref{fig:exp}. This quantity relates directly with the effective mechanical properties of a specific choice of unit cell geometry; when under fracture processes, the lattices will carry load up to failure and their effective stiffnesses are found to scale with their cross-sectional area of the strut in which are made, \emph{i.e.} proportional to $\tilde{\ell}^2$. Moreover, it can be shown that stretching and bending dominated unit cells result into lattices with effective stiffnesses scaling respectively with $\tilde{\ell}^2$ and $\tilde{\ell}^4$~\citep{Gibson1997}. Physically, this means that in the stretching dominating regime tensile stresses in strut are prevalent, whereas coupled tensile-compressive stresses are dominant for bending dominated unit cells. Therefore, we may write that the lattice interface effective stiffnesses of the choices of unit cells are given the following expressions:
\begin{subequations}
\label{eq:E_effective}
        \begin{align}
		&E_{\mathrm{|}}=2\pi\tilde{\ell}^2E^*,\label{eq:E_pil}\\        
		&E_\mathrm{\triangle}=C_{45}\tilde{\ell}^2E^*,\label{eq:E_str}\\
		&E_\mathrm{\hexagon}=C_{60}\tilde{\ell}^4E^*, \label{eq:E_ben}
        \end{align}
\end{subequations}
where $E_{i}$ represents the effective stiffness of each lattice, the label ${i}\in\{\mathrm{|},\mathrm{\triangle},\mathrm{\hexagon}\}$ indicates which unit cell makes up the lattice ($\mathrm{|}=\,$pillars, $\mathrm{\triangle}=\,$triangular and $\mathrm{\hexagon}=\,$hexagonal), and $E^*$ stands for the stiffness of the base material. The geometrical constants $C_{45}=\pi\left(3+\sqrt{2}\right)/7$ and $C_{60}=32\pi\sqrt{3}$ are obtained for the $45^\circ$ triangular horizontal lattice and $60^\circ$ regular honeycomb~\citep{Gibson1997}. \com{Notice that, from the theoretical perspective, using 3D formulation would merely result in modified set of constants $E_i$.} Furthermore, we refer to the normalisation with respect to the base material of the interface as $\tilde{E}_{i}\equiv E_{i}/E^*$. The three lattice interfaces are considered as homogeneous brittle structures carrying uni-axial loading. Therefore, an upper and a lower bound for the localised normal strain is set for each type of geometry. The upper (tensile) bound describes the reach to tensile failure stress $\sigma_{\!f}$ at an individual strut, and the lower (compressive) bound corresponds to reaching the Euler buckling load at an individual strut. This last failure mechanism forms an augmented version of the Cohesive Zone Model (CZM)~\citep{Barenblatt1962,Dugdale1960,Hillerborg1976} incorporating compressive regime which characterises the lattice bondline. Specifically for the pillar unit cell, the tensile fracture external stress is $2\pi\tilde{\ell}\sigma_{\!f}$ and the buckling (assuming fixed boundary conditions) external stress is $2\pi^3\tilde{\ell}^4E^*$, since the maximum internal force and total external force scale with $1/2$. Hence, the bounds for tension $(t)$ and compression $(c)$ are respectively given as follows:
\begin{subequations}
        \begin{align}
		\varepsilon_\mathrm{|}^{(t)}&=\frac{\sigma_{\!f}}{E^*}\\
		\varepsilon_\mathrm{|}^{(c)}&=-\pi^2\tilde{\ell}^2.
         \end{align}
         \label{eq.pillars}
\end{subequations}
For the triangular unit cells, the fracture external stress is $\sqrt{2}\pi\tilde{\ell}\sigma_{\!f}$ and, assuming pinned boundary conditions, the buckling external stress is $\sqrt{2}\pi^3\tilde{\ell}^4E^*/2$, since the maximum internal force and total external force scale with $2^{-1/2}$. This results in the following expressions for the bounds in the critical stain:
\begin{subequations}
        \begin{align}
		\varepsilon_\mathrm{\triangle}^{(t)}&=\frac{\sqrt{2}\pi}{C_{45}}\frac{\sigma_{\!f}}{E^*}\\
		\varepsilon_\mathrm{\triangle}^{(c)}&=-\frac{\sqrt{2}\pi^3}{2C_{45}}\tilde{\ell}^2.
         \end{align}
         \label{eq:Tri}
\end{subequations}
For the last example, the bending dominated hexagonal unit cell, failure is achieved when the ultimate stress is reached as bending stress. The corresponding bending moment at failure in the strut is $\pi\sigma_{\!f} r^3/2$, thus providing a stress in the unit cell given by $16\pi\sigma_{\!f}\tilde{\ell}^3/3$. Because the unit cell is bending dominated, its response is symmetric in tension and compression. Hence, we arrive at the following bounds:
\begin{subequations}
        \begin{align}
		\varepsilon_\mathrm{\hexagon}^{(t)}&=\frac{16\pi}{3C_{60}}\frac{\sigma_{\!f}}{E^*}\tilde{\ell}^{-1}\\
		\varepsilon_\mathrm{\hexagon}^{(c)}&=-\varepsilon_\mathrm{\hexagon}^{(t)}.
         \end{align}
         \label{eq:hex}
\end{subequations}
This formulation of the bounds uncovers a dependence between the lattice failure mode and the geometry. More specifically, it will later determine the critical loads of the stretching dominated cells relate to the geometry at the compressive regime, while the tensile fracture load is geometrically independent. Regarding bending dominated cells, the dependence exists in both fracture modes but appears at the denominator.
\section{ Cohesive Zone Model for thick bondlines}
\label{analytical}
The present problem can be treated as a particular case of an adhesive joint with discontinuous, architected interface region, \emph{i.e.} the bondline. We consider a well known, and reliable, Double Cantilever Beam (DCB) configuration under fracture loading~\citep{kanninen}, outlining some differences between homogeneous solids and the lattice (metamaterial) interfaces. For instance, such formulation does not capture details of local effects experienced by joints with thick bondlines and due to the crack front sharpness, as $a_0\sim h_i$, or bi-material edges~\citep{mohammed2000cohesive,LopesFernandes2021Stiff}. However, it can be found suitable here, as lattice interfaces can be used to mitigate such local effects. In addition, for fracture of lattice structures, presence of sharp cracks can be debated. For the sharp crack tip is associated with existence of characteristic material length scale parameter $l_c\propto \alpha K_{c}^2/ \sigma_y^2$, where $K_c$ is the relevant, critical stress intensity factor, $\sigma_y$ is the failure stress (\emph{i.e.} yield stress for ductile materials, or fracture stress for brittle), and $\alpha$ is a constant dependent on geometry of the problem and stress state at the crack tip. The spurious point originates from the fact that, even for brittle materials like polymethylmethacrylate (PMMA or plexiglass), $l_c$ equates to \emph{ca.} $100$ micrometers. For popular materials used within the additive manufacturing, such as stainless steels, aluminium and number of structural polymers, $l_c$ exceeds several millimetres. With lattice structure unit cell walls or trusses measuring usually nano-to-millimetres, sharp crack assumption cannot be met, and alternative way for studying progressive failure should be proposed.

\begin{figure}
    \centering
    \includegraphics[width=.5\textwidth]{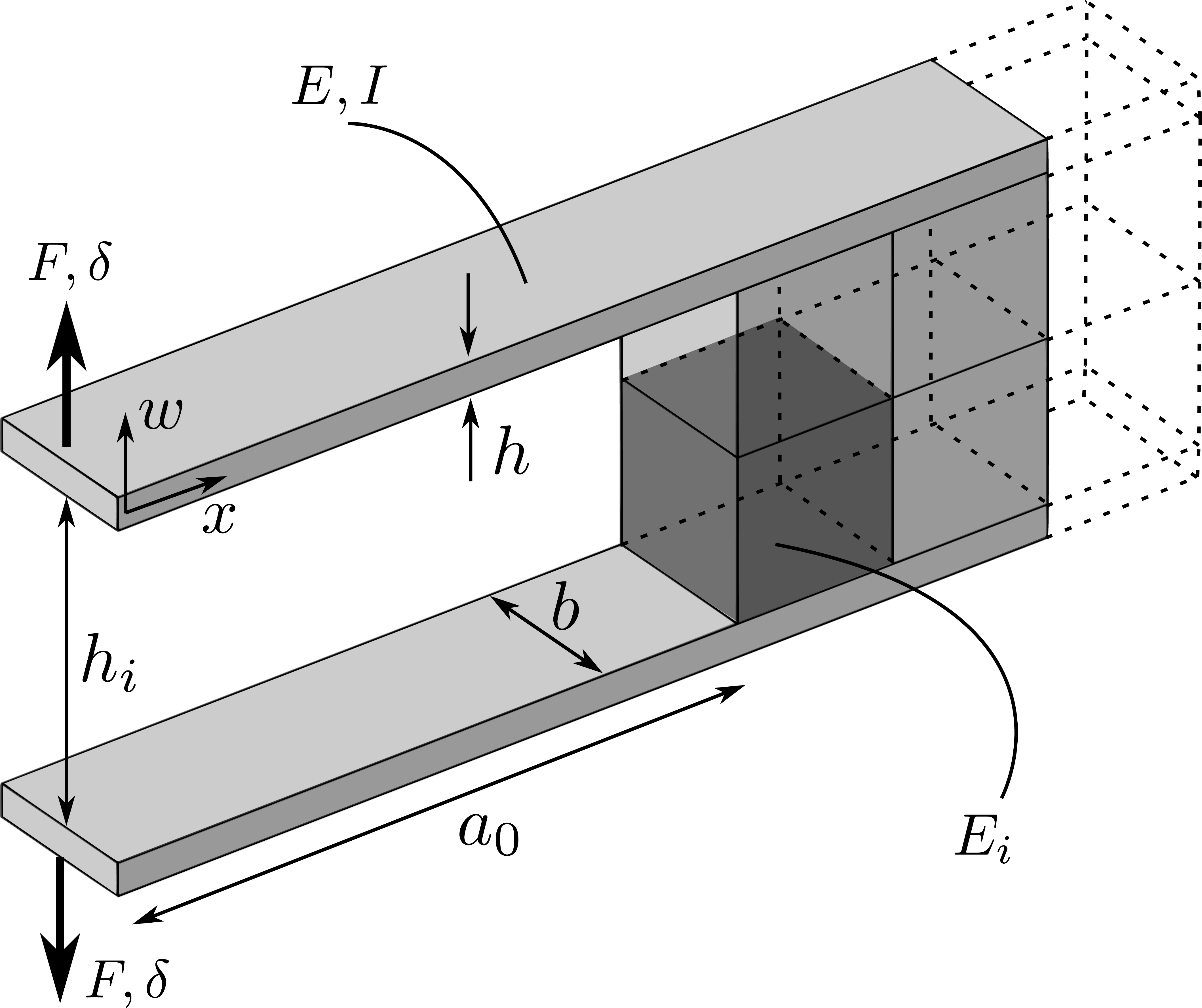}
    \caption{Schematic of the followed Double Cantilever Beam setup.}
    \label{fig:dcb_setup}
\end{figure}

In order to capture interaction between the lattice interface and joined material the DCB configuration, as represented in Fig.~\ref{fig:dcb_setup} is adopted.  
Mechanical behaviour of such material system originates effectively from a bonded structure under fracture loading, and it can be seen and modelled as a generic beam on elastic foundation~\citep{dillard2018review}. The following equilibrium equation gives us the minimiser for the elastic strain energy of such system:
\begin{equation}
	E\,I\frac{\mathrm{d}^4w(x)}{\mathrm{d}x^4}=p(x)+q(x),
	\label{eq:winkler}
\end{equation}
where $E$ and $I=bh^3/12$ are the Young's modulus and moment of area of the beam, respectively. The relevant geometric parameters associated to the beam are its width $b$ and thickness $h$, assuming a rectangular cross section. On the right hand side of Eq.~\eqref{eq:winkler}, $p(x)$ and $q(x)$ are forces per unit-length, respectively, representing the external applied load and the stresses due to the elastic foundation. Since $q(x)$ yields a restoring force, and assuming the elastic foundation to be linear, we have that $q(x)=-k(x)w(x)$, where $k(x)$ is the elastic foundation stiffness (force per unit length square). This stiffness is expressed as follows,
\begin{equation}
	k(x)=2\frac{b}{h_{{i}}} E_{{i}}H(x-a),
	\label{eq:winkler2}
\end{equation}
where $a$ is the crack length, $E_{{i}}$ is the effective stiffness of each choice of lattice, given by Eq.~\eqref{eq:E_effective}, $h_{{i}}$ is the thickness of each interface, and $H(x-a)$ is the Heaviside step function determining the location of the crack front---notice that the domain in, $0\le x<\infty$, is such that for $0\le x<a$ the beam is debonded and the foundation only acts for $a\le x<\infty$. Finally, the applied load is simple shear at the boundary, which is expressed by $p(x)=F\delta(x)$, where $\delta(x)$ is the Dirac delta function necessary to localise the applied load $F$ (units of force) acting on the boundary $x=0$. Before proceeding with the solution to the Eq.~\eqref{eq:winkler}, we acknowledge the fact that this equation gives rise to a characteristic wave-length for the region over which crack front stress can be distributed, which is here derived from the length scales in the problem as well as stiffnesses' ratio:
\begin{equation}
	\lambda_{i}=\lambda_0\,\left(\frac{\tilde{h}_{i}}{\tilde{E}_i/\bar{E}}\right)^{1/4},
	\label{eq:winkler3}
\end{equation}
where $\lambda_0\equiv h/6^{1/4}$ is the resulting wave-length when the interface thickness is negligible~\citep{kanninen}, $\tilde{h}_{i}\equiv h_i/h$ is the thickness ratio used as one of our design parameters, and $\bar{E}\equiv E/E^*$ is the stiffness' ratio of the beam and the lattice's base material. The above characteristic wave-length, along with the definition of dimensionless quantities\footnote{Notice that when writing the Dirac delta function in the dimensionless form, we have that $\delta(x)=\delta(\tilde{x}\lambda_{i})=\delta(\tilde{x})/\lambda_{i}$.}, gives us the following dimensionless form of Eq.~\eqref{eq:winkler}: 
\begin{equation}
	\frac{\mathrm{d}^4\tilde{w}(\tilde{x})}{\mathrm{d}\tilde{x}^4}=\tilde{F}_{i}\delta(\tilde{x})-4H(\tilde{x}-\tilde{a})\tilde{w}(\tilde{x}),\quad 0\le \tilde{x}<\infty,
	\label{eq:winkler4}
\end{equation}
where $\tilde{w}\equiv w/h_{{i}}$, $\tilde{x}\equiv x/\lambda_{i}$, $\tilde{a}\equiv a/\lambda_{i}$, and $\tilde{F}_{i}\equiv \left[\lambda_{i}^3/\left(E Ih_{{i}}\right)\right]F$. Ensuring that the edge of the beam is moment free, \emph{i.e.} $\mathrm{d}^2\tilde{w}(0)/\mathrm{d}\tilde{x}^2=0$, as well as that the solutions are at least $\mathcal{C}^3$ at $\tilde{x}=\tilde{a}$---thus ensuring continuity of the solution, slop, moment, and shear---we find the following piece-wise solution
\begin{equation}
	\tilde{w}(\tilde{x})=\frac{\tilde{F}_{i}}{2}\times
	\begin{cases}
	    \begin{aligned}
			\frac{1}{3}\left[1+\tilde{x}^3+2\left(1+\tilde{a}\right)^3-3\tilde{x}\left(1+\tilde{a}\right)^2\right], \tilde{x}<\tilde{a}\\
			e^{\tilde{a}-\tilde{x}}\left[\left(1+\tilde{a}\right)\cos\left(\tilde{a}-\tilde{x}\right)+\tilde{a}\sin\left(\tilde{a}-\tilde{x}\right)\right), \tilde{x}\ge\tilde{a}.	
        \end{aligned}
    \end{cases}
	\label{eq:winkler_sol}
\end{equation}

Contrary to solid interfaces for which in the present DCB configuration compressive failure will never be observed, confined lattice interfaces will likely be sensitive to the direction of loading~\citep{heide2020mechanics}. In order to highlight these differences, we shall here define a tension-compression amplification factor. Let us look at the first two extrema of Eq.~\eqref{eq:winkler_sol} in the interface region ($\tilde{x}\geq \tilde{a}$), which are found at $\tilde{x}=\tilde{a}$ for the maximum and at $\tilde{x}=\tilde{a}+\pi-\tan^{-1}\left(1+2\tilde{a}\right)$ for the minimum. The corresponding deflection values are found:
\begin{subequations}
    \label{eq:maxmin}
        \begin{align}
        \tilde{w}_{\max}&=\frac{\tilde{F}_{i}}{2}\left(1+\tilde{a}\right)
        \label{eq:max}
        \\
        \tilde{w}_{\min}&=-\frac{\tilde{F}_{i}}{2}\sqrt{\frac{1}{2}+\tilde{a}\left(1+\tilde{a}\right)}e^{\arctan\left(1+2\tilde{a}\right)-\pi}.
        \label{eq:min}
        \end{align}
\end{subequations}
Therefore, a tension-compression amplification factor is can be defined as follows:
\begin{equation}
	\hat{w}\equiv\left|\frac{\tilde{w}_{\max}}{\tilde{w}_{\min}}\right|=\frac{1+\tilde{a}}{\sqrt{\frac{1}{2}+\tilde{a}\left(1+\tilde{a}\right)}}e^{\pi-\arctan(1+2\tilde{a})},
	\label{eq:minmaxratio}
\end{equation}
where the bounds are given by $e^{\pi/2}<\hat{w}<\sqrt{2} e^{3\pi/4}$.

We now propose to solve a modified version of Eq.~\eqref{eq:winkler} by including a CZM~\citep{Barenblatt1962,Dugdale1960,Hillerborg1976}  in both
tensile and compressive loading directions. More specifically, we distinguish between three possible regimes of the interface behaviour. In the elastic regime of the interface the stresses are simply proportional to the beam deflection, \emph{i.e.} $\tilde{\sigma}\equiv\sigma/E^*=2H(\tilde{x}-\tilde{a})\tilde{E}_{i}^{(\cdot)}\tilde{w}$. 
When the largest negative deflection is less than some critical value $\varepsilon_{i}^{(c)}$, it is assumed that the interface fails instantaneously (\emph{i.e.} through brittle failure) and no longer carries any stress. Hence, this compressive failure scenario is avoided when $2\tilde{w}>\varepsilon_{i}^{(c)}$. When the largest positive deflection is larger than critical value $\varepsilon_{i}^{(t)}$, the stress carried by the interface is locally zero and the interface fails gradually. All of these facts are summarised by the following conditional statement:
\begin{equation}
	\tilde{\sigma}=H(\tilde{x}-\tilde{a})
	\left\{\begin{array}{llrcl}
		0&,&&\!\!\!\!2\tilde{w}\!\!\!\!&<\varepsilon_{i}^{(c)}\\
		2\tilde{E}_i\tilde{w}&,&\varepsilon_{i}^{(c)}\leq&\!\!\!\!2\tilde{w}\!\!\!\!&<\varepsilon_{i}^{(t)}\\
		0&,&&\!\!\!\!2\tilde{w}\!\!\!\!&\geq \varepsilon_{i}^{(t)},
	\end{array}\right.
	\label{eq:CZM}
\end{equation}
Crack initiation due to tension happens when the upper bound of the unit cell strain is reached, $2\tilde{w}_{\max}=\varepsilon_{i}^{(t)}$, while failure due to compressive loads is observed when the lower bound of the unit cell strain is reached, $2\tilde{w}_{\min}=\varepsilon_{i}^{(t)}$. From Eq.~\eqref{eq:maxmin}, we may write the critical reaction force due to tension and compression of each unit cell, respectively:
\begin{subequations}
\label{eq:collapse0}
\begin{gather}
    \tilde{F}_{i}^{(t)}=\frac{1}{1+\tilde{a}}\varepsilon_{i}^{(t)},\\
    \tilde{F}_{i}^{(c)}=-\frac{e^{\pi-\arctan(1+2\tilde{a})}}{\sqrt{\frac{1}{2}+\tilde{a}\left(1+\tilde{a}\right)}}\varepsilon_{i}^{(c)}.
\end{gather}
\end{subequations}
Similarly to Eq.~\eqref{eq:minmaxratio}, a critical strain factor can be defined as: $\hat{\varepsilon}=|\varepsilon_{i}^{(t)}/\varepsilon_{i}^{(c)}|$. Since the tension-compression amplification factor, $\hat{w}$, is independent of the normalised force, the failure mechanism is instead imposed in comparison with $\hat{\varepsilon}$. This follows the failure criterion due to tension, which happens when $\hat{w}>\hat{\varepsilon}$, and the one due to compression, when $\hat{w}\leq\hat{\varepsilon}$.

\section{Numerical modelling of architected interface failure}
\label{Section:FE}

Since both stretching and bending dominated cells are studied, the failure modes of the interface's individual struts are predominantly subjected to axial stresses and bending moments, respectively. Therefore, in our Finite Element Method (FEM) setup, we model all of these elements, in either stretching or bending dominated cells, as Euler-Bernoulli beam elements. We also assume that the interface's base material properties is such that its behaviour is brittle. Hence, all the failure cases were considered instantaneous and the behaviour of the material was modelled as linear elastic
until failure occurs. 

For stretching dominated unit cells, we assume failure under tension occurs when the ultimate tensile stress is applied to the struts. Therefore, a maximum tensile strain is obtained by $\varepsilon_u^{(t)}=\sigma_{\!f}/E^{*}$.
On the other hand, when in compression, the strut is considered slender thus buckling occurs before reaching the maximum compressive stress. The critical buckling load, considering built-in ends, is given as $P_c=\pi^3 E^{*}r^4/s^2$. Therefore, the maximum compressive strain for a stretching dominated strut with circular cross section, is given by $\varepsilon_u^{(c)}=-P_c/(E^{*}A)=-\pi^2\tilde{\ell}^2$. Lastly, a bending dominated strut may fail under normal stresses during bending. Hence, there is only one possible failure mode, which is given by a maximum normal strain that a bending dominated strut
may carry. This case is given by the ultimate strain $\varepsilon_{u}^{(t)}=-\varepsilon_u^{(c)}=\sigma_{\!f}/E^{*}$. \com{Notice that such statement is similar to that of eq. (\ref{eq.pillars}), with the difference that in the analytical formulation of failure strains, different unit cells result in distintic geometrical pre-factors, \emph{viz.} eqs. (\ref{eq:Tri}) and (\ref{eq:hex})---such pre-factors naturally come from the full geometric representation within the numerical framework.}

The FEM analyses were carried using COMSOL Multiphysics (version 5.5). The brittle failure of the struts was implemented manually in the definition of the Young's modulus of the lattice material~\citep{vermeltfoort2013}. In specific, the Young's modulus was defined as a function of the
appropriate strain for each case, resulting into the following composed function:
\begin{equation}
	E(\varepsilon)=\begin{cases}\begin{aligned}
			0,&& \varepsilon&<\varepsilon_u^{(c)}\\
			E^*,&& \varepsilon_u^{(c)}\leq\varepsilon&<\varepsilon_u^{(t)}\\
			0,&& \varepsilon_c^{(t)}<\varepsilon&
	\end{aligned}\end{cases}
	\label{eq:modulus}
\end{equation}
The proposed expression in Eq.~\eqref{eq:modulus} brings about the following problem in the FEM implementation: the value of the Young's Modulus can neither be zero nor discontinuous. To overcome this issue, the function is smoothed using COMSOL's built in step function ensuring $\mathcal{C}^2$ continuity and introducing a transition zone with size a fraction of the respective ultimate strain value. Moreover, the Young's modulus of the failed lattice was taken between $10^{-4}E^*$ and $10^{-3}E^*$ instead of zero---this is done for regularisation purposes, which maintains the elements after fracture carrying very low stress values.

\begin{figure}
\centering
\subfloat[]{\includegraphics[width=\textwidth]{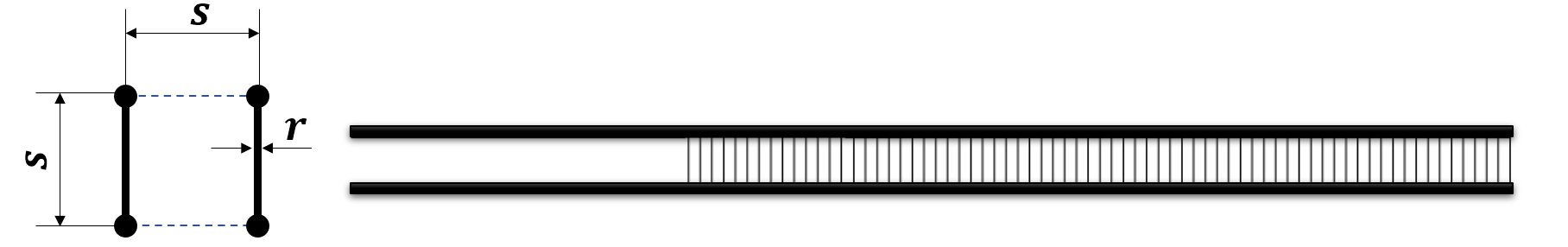}}\\
\hfill
\subfloat[]{\includegraphics[width=\textwidth]{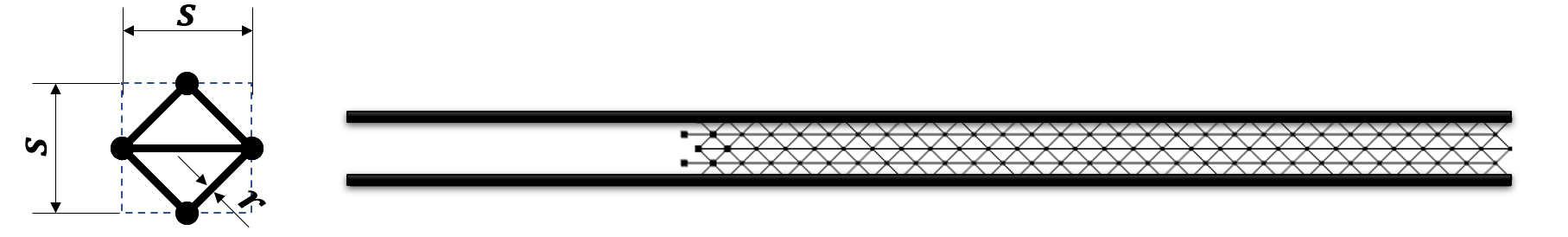}}\\
\hfill
\subfloat[]{\includegraphics[width=\textwidth]{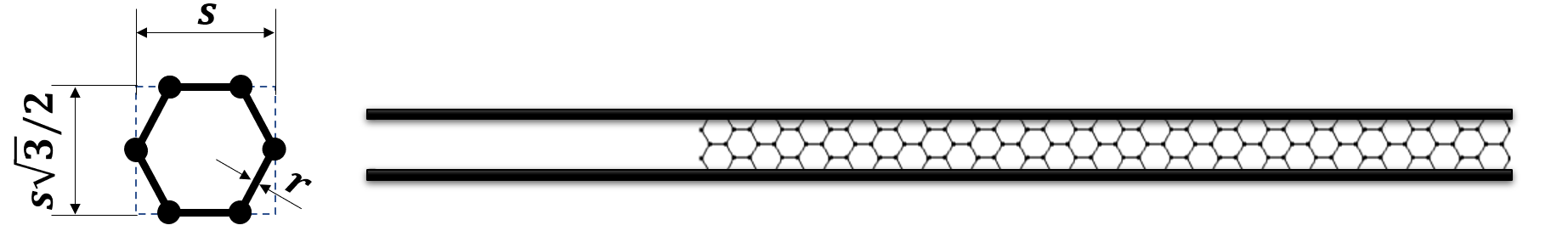}}
\caption{(a) Stretching dominated pillar interface. (b) Stretching dominated triangular interface. (c) Bending dominated hexagonal interface. Schematic representation of 2D Finite Element setup with different unit cells used for interfaces. The entire geometry, both the composite adherents and the lattice interface are discretised using beam elements.}
\label{fig:Schematic_FE}
\end{figure}

The geometry of the model, shown in Fig.~\ref{fig:Schematic_FE}, consists of a sufficiently long DCB setup, divided into free and bonded parts, and an array of the modelled lattice structure. Symmetry along the centre-line, which cuts through the longitudinal direction of the geometry, was utilised in order to model only the upper half of the DCB setup---except for the pillar in order to achieve a single element per pillar and avoid numerical instabilities regarding the horizontal displacements of the vertices positioned on the symmetry line. An Euler-Bernoulli beam type elements were used throughout, where the geometric differences were input in the cross section data. More specifically, the elements forming the cantilever beam are assumed to have a rectangular cross-section, $bh$, while the elements forming all the different types of unit cells' interface have a cylindrical cross section of radius $r=\tilde{\ell}s$---in the simulations we take approximately in the range $\tilde{\ell}\in[0.001,0.1]$. The selected beam dimensionless thickness took values $\tilde{h}=\{0.1,1,10\}$ (the initial crack length, $a_0$ was $15h_i$, $10h_i$ and $5h_i$, for each respective value of $\tilde{h}$). The width of the beam is chosen to be equal to the unit cell length and two unit cells are placed through the interfacial thickness. Hence, $b=s=h_i/2$ for pillar and stretching dominated cells and $b=s=h_i/\sqrt{3}$ for the bending dominated cell. Moreover, the beam stiffness $E=100GPa$ and lattice stiffness $E^*=5GPa$ were taken across all the simulations. The boundary conditions were setup such that the free end of the beam is mounted to a rigid connector which ensures free rotation alongside with prescribed displacements. The symmetry line is also a boundary condition constraining the vertical displacement of the selected points. A stationary solver is used in which, pseudo-time is included as a continuation parameter to model the moving action of a tensile machine. Every strut in the model is represented by only one beam type element in order to increase computational stability and eliminates the occurrence of local phenomena during failure. 

\section{Discussion of the results}
Having established the model, in this section we gather the results of the analytical and numerical analyses. Initially, the quantities of interest are the normal displacement fields from Eq.~\eqref{eq:winkler_sol}, for all the interfaces shown in Fig.~\ref{fig:Schematic_FE}, and their respective fracture onset forces given by Eq.~\eqref{eq:collapse0}. Furthermore, we gain insight towards achieving an unified designing principle for metamaterial interfaces by looking at the effect of the fracture onset forces with respect to the materials' density as well as the characteristic wave-length, from Eq.~\eqref{eq:winkler3}. This insight is presented together with the properties for joints used in bulk interfaces, such as face-core debonding~\citep{Prasad1994debonding}, stiff~\citep{LopesFernandes2021Stiff} and soft~\citep{Cabello2016Soft} epoxy adhesives.

\subsection{Behaviour under load}

\begin{figure}[h!]
\centering
\subfloat[]{\includegraphics[width=0.5\textwidth]{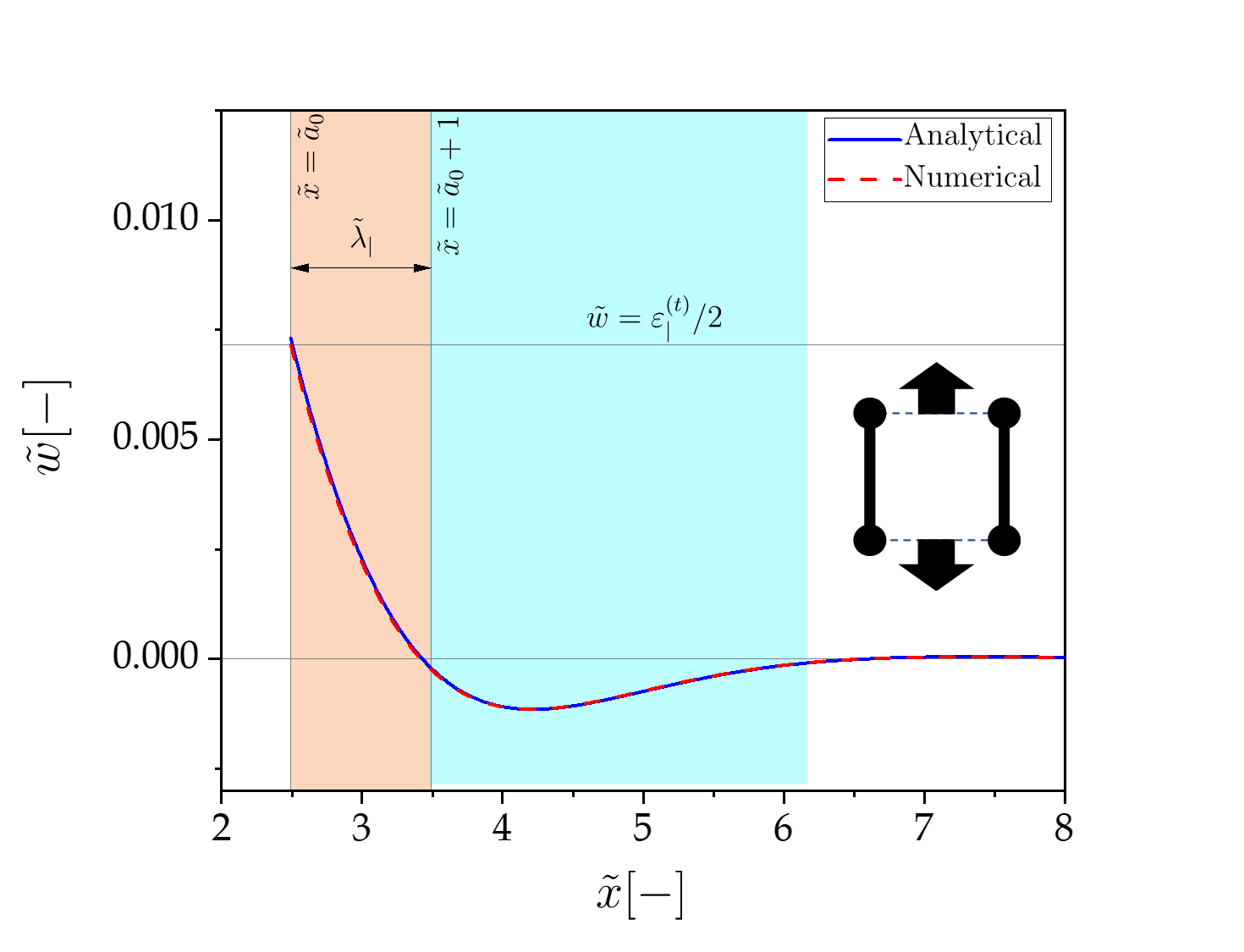}}
\subfloat[]{\includegraphics[width=0.49\textwidth]{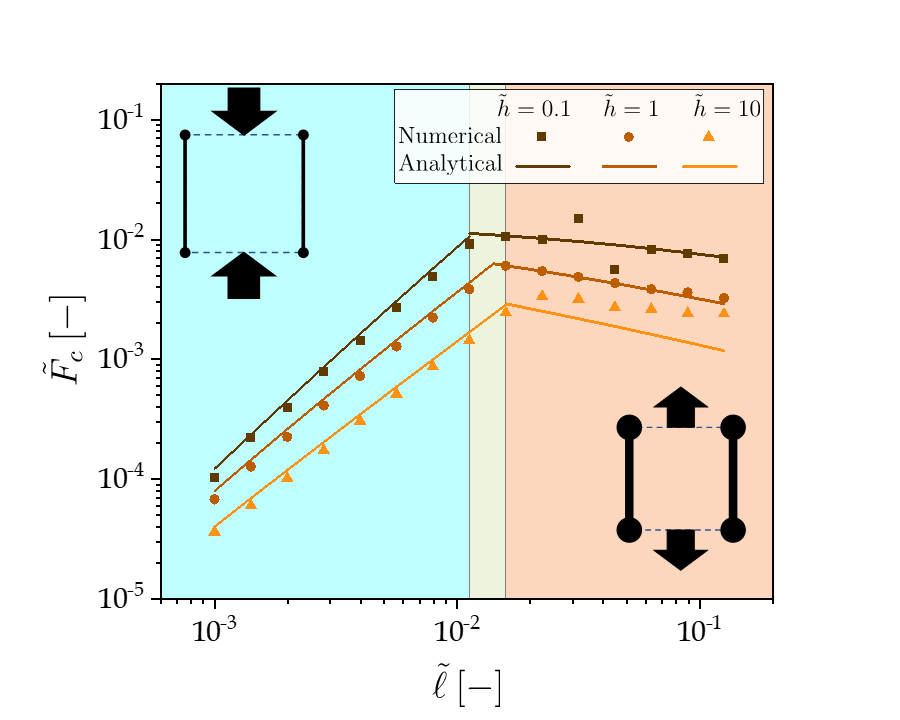}}
\caption{(a) Example, $\tilde{\ell}=4.5\times10^{-2}$ and $\tilde{h}=1$ of the normalised deflection field for the pillar interface under tensile failure mode. Both models indicated similar deflection at failure $\tilde{w}=\varepsilon_|^{(t)}$ and length of the process zone $\tilde{\lambda}_{|}$ and deflection field distribution - red background for tensile and blue for compressive deflections. (b) Failure map for pillar geometry where both power-law are visible as well as the effects of the interface thickness. Two regions are found: In low aspect ratios, the compressive displacement related to the buckling is reached first and the critical force recorded is related to the buckling load. In high aspect ratios, the tensile displacement related to the yielding of the pillars is reached first and the critical force recorded is related to the yield stress of the interface material.}
\label{fig:Pillars}
\end{figure}
\begin{figure}
\centering
\subfloat[]{\includegraphics[width=0.49\textwidth]{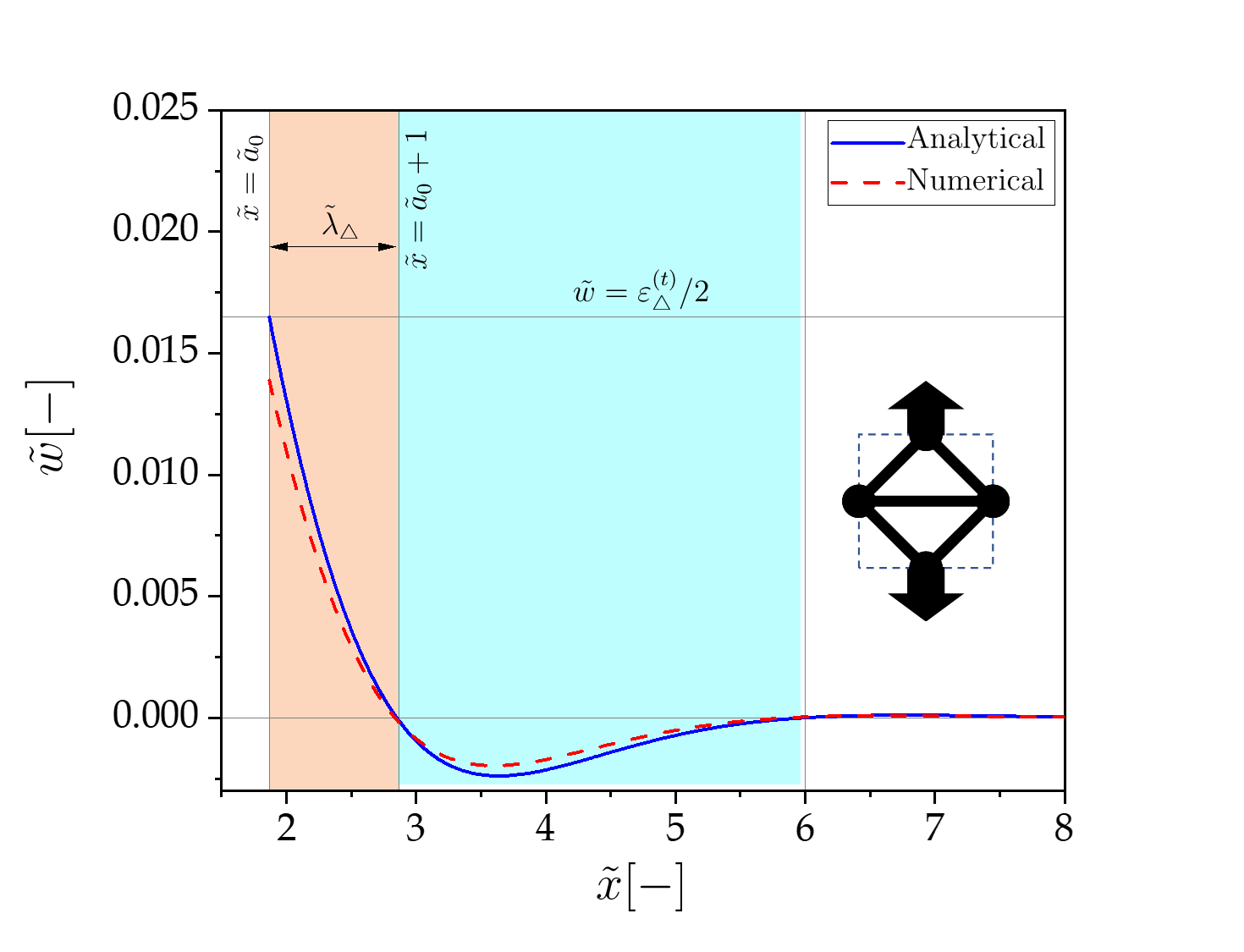}}
\subfloat[]{\includegraphics[width=0.49\textwidth]{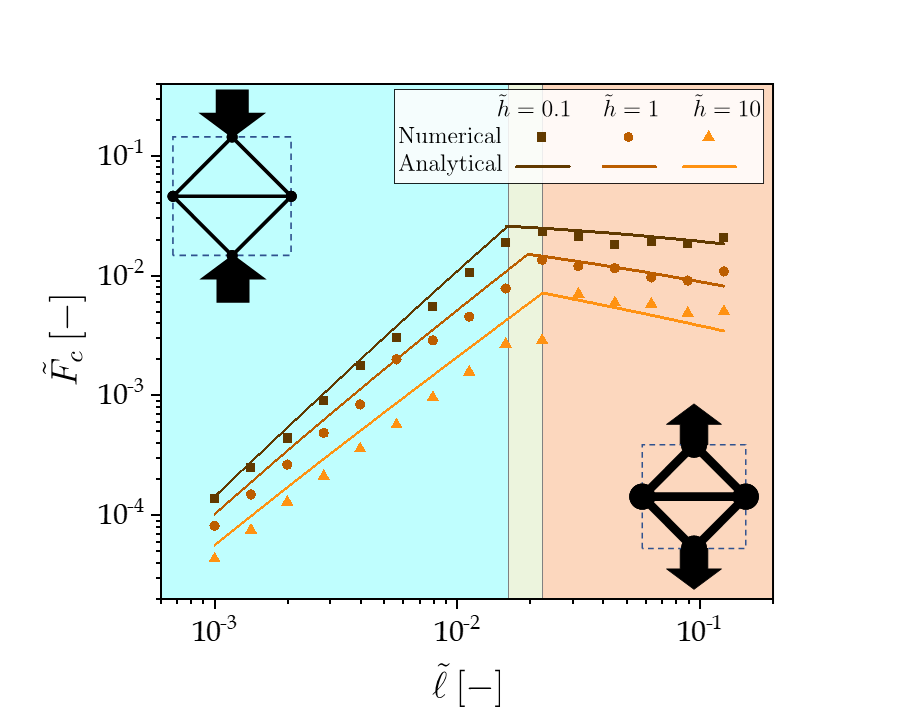}}
\caption{(a) Example, $\tilde{\ell}=4.5\times10^{-2}$ and $\tilde{h}=1$ of the normalised deflection field for the octet-like interface under tensile failure mode. Both models indicated similar deflection at failure $\tilde{w}=\varepsilon_{\triangle}^{(t)}$ and length of the process zone $\tilde{\lambda}_{\triangle}$ and deflection field distribution - red background for tensile and blue for compressive deflections.)
Failure map for octet truss geometry. Two regions are found: In low aspect ratios, the compressive displacement related to the buckling is reached first and the critical force recorded is related to the buckling load. In high aspect ratios, the tensile displacement related to the yielding of individual struts of the unit cell is reached first and the critical force recorded is related to the yield stress of the interface material.}
\label{fig:Octet}
\end{figure}
\begin{figure}
\centering
\subfloat[]{\includegraphics[width=0.49\textwidth]{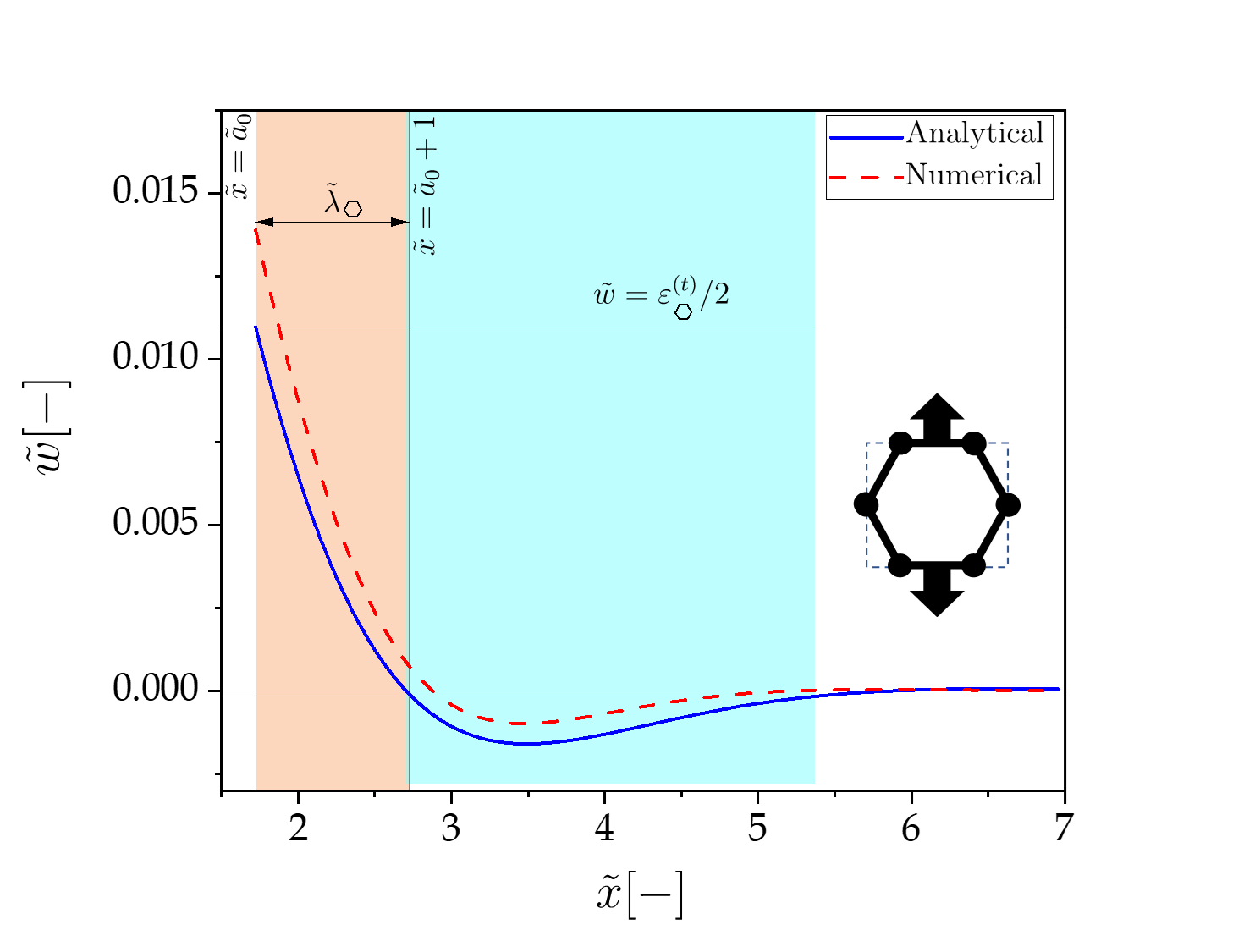}}
\subfloat[]{\includegraphics[width=0.49\textwidth]{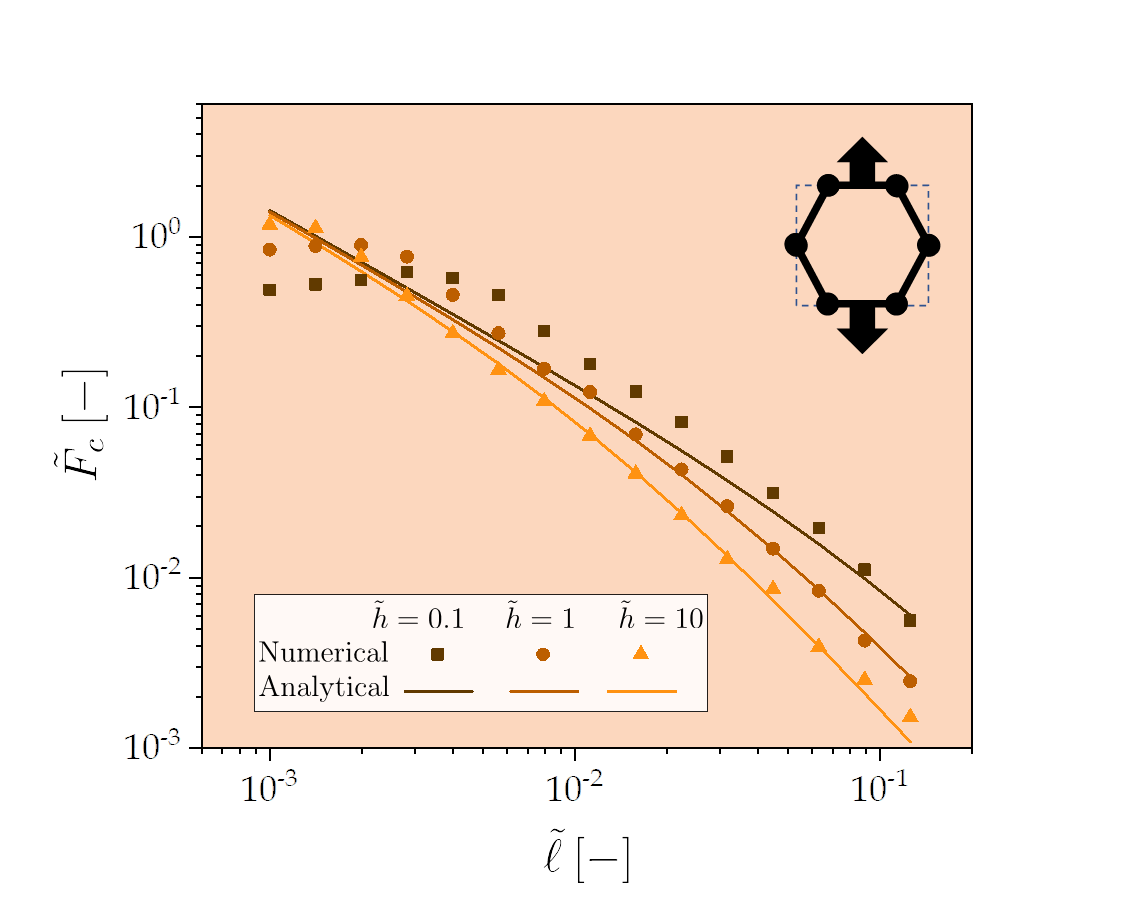}}
\caption{(a) Example, $\tilde{\ell}=6.3\times10^{-2}$ and $\tilde{h}=1$ of the normalised deflection field for the bending dominated interface under tensile failure mode. Both models indicated similar deflection at failure $\tilde{w}=\varepsilon_{\hexagon}^{(t)}$ and length of the process zone $\tilde{\lambda}_{\hexagon}$ and deflection field distribution - red background for tensile and blue for compressive deflections. (b) Failure map for bending dominated geometry. Note, that for the bending dominated unit cell, only a single, tensile, failure mode exists.}
\label{fig:Kelvin}
\end{figure}

In the Figs.~\ref{fig:Pillars}, \ref{fig:Octet}, and \ref{fig:Kelvin} we discuss the effects the choice of unit cell topology has on the type of failure\com{, which is measured through the normalised critical force $\tilde{F}_{c}\equiv \left[\lambda_{i}^3/\left(E Ih_{{i}}\right)\right]F_{c}$ of the fracture onset---notice that this normalisation naturally encodes the characteristic process zone length $\lambda_i$ and material properties}. Firstly, in all cases presented in Figs.~\ref{fig:Pillars}-\ref{fig:Octet}-\ref{fig:Kelvin} part (a), we observe that the deflection of the lever arm in the DCB configuration manifests two regions: (i) $\tilde{a}_0\le\tilde{x}<1+\tilde{a}_0$ \com{(given our choice of normalisation by $\lambda_i$, we explicitly choose to show the size $\tilde{\lambda}_i=1$ in the figures for a clear graphic representation of the process zone)}, where the foundation is subject to tension; and (ii) $\tilde{x}\ge1+\tilde{a}_0$, for when the metamaterial interface is under compression.
Notice that the largest deflection is set by $\varepsilon_i^{(t)}$. Now let us turn our attention to classifying the different types of phases. This is accomplished by taking a designers approach to the problem at the element scale, \emph{i.e.} struts, where we look at the effect on the fracture onset forces from the inverse of the slenderness parameter, namely the aspect ratio $\tilde{\ell}$. Both stretching dominated geometries, pillars shown in Fig.~\ref{fig:Pillars}-(b) and the triangular unit cells Fig.~\ref{fig:Octet}-(b), show a qualitative similar behaviour. For low aspect ratio $\tilde{\ell}$, the struts are slender and the structure is more susceptible to failure under collapse; on the other hand, at higher values of $\tilde{\ell}$, \emph{i.e.} stubbier elements, the power-law changes and the onset of failure occurs in tension. The failure map results for the hexagonal lattice are shown in Fig.~\ref{fig:Kelvin}. As expected, the hexagonal unit cells being bending dominated, will only fail in tension since any compression will be uptake by moments on the hinges.

\subsection{Comparisons and tools for designers}

\begin{figure}
    \centering
    \includegraphics[width=.5\textwidth]{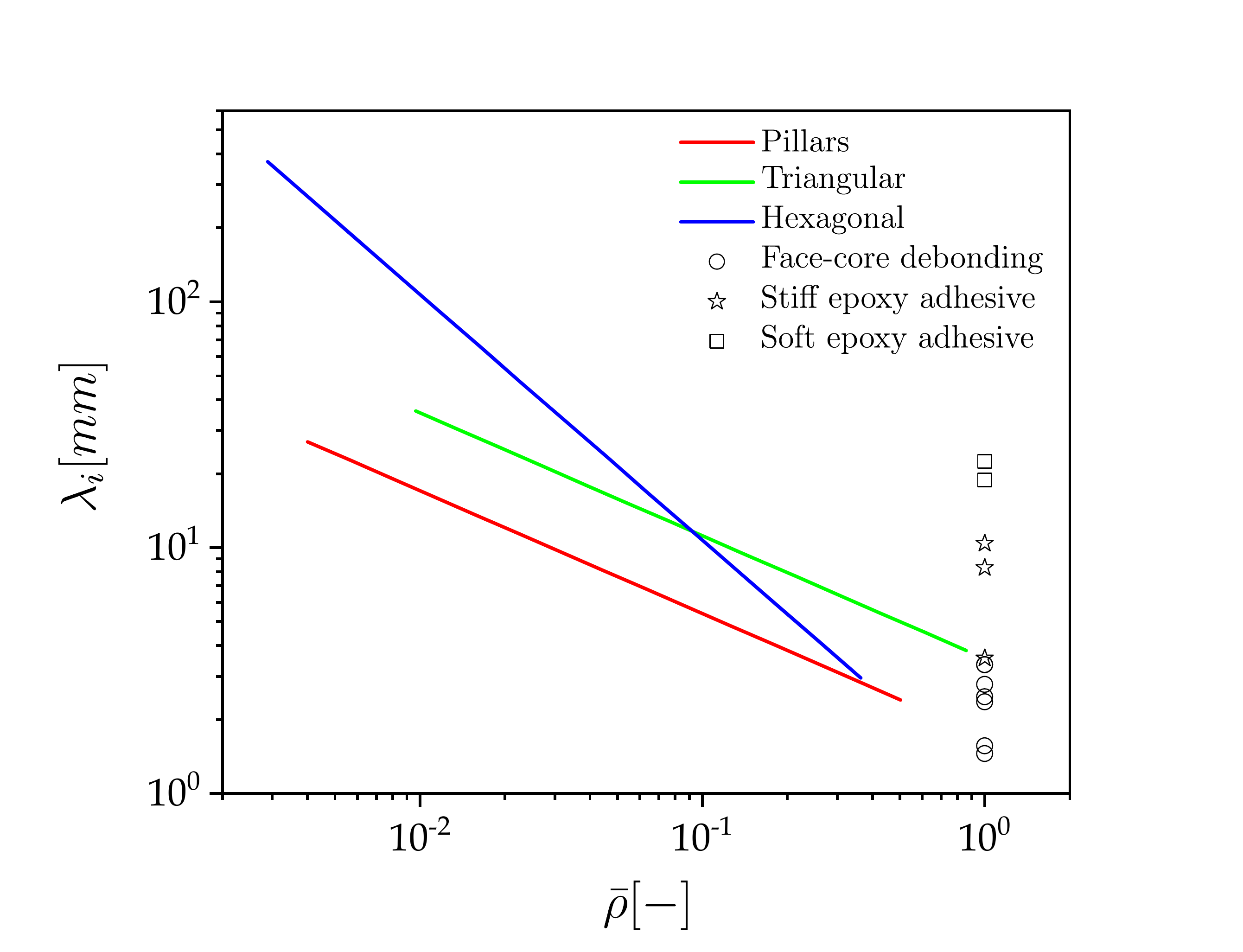}
    \caption{Analytical wavelength with respect to mid-plane areal density for $\tilde{h}=1$. Empirical results for sandwich material core debonding ($\bigcirc$), stiff epoxy adhesive ($\openbigstar$) and toughened epoxy adhesive ($\square$) are shown accordingly~\citep{Prasad1994debonding,LopesFernandes2021Stiff,Cabello2016Soft}.}
    \label{fig:lambda}
\end{figure}

One of the characteristic feature of materials with confined interfaces is existence of the dominating length scale $\lambda_i$, which describes interaction between the joined material and the interface. In Fig.~\ref{fig:lambda}, the relation between the $\lambda_i$ and $\bar{\rho}$ is depicted. In particular, $\bar{\rho}$ is defined as the fraction area occupied by the lattice microstructure over the area of the unit cell, i.e. $\bar\rho_\mathrm{|}=4\tilde{\ell}$, $\bar\rho_\mathrm{\triangle}=4(1+\sqrt{2})\tilde{\ell}$, $\bar\rho_\mathrm{\hexagon}=5\tilde{\ell}/\sqrt{3}$---notice that, for consistency, we account for the third dimension by computing the volume fractions, despite the problem being formulated in the state of plane stress.

For comparison, a few empirical results from the literature studying behaviour of popular bulk interfaces are also presented. We look at the cases of sandwich material core debonding~\citep{Prasad1994debonding}, stiff epoxy adhesive~\citep{LopesFernandes2021Stiff} and toughened epoxy adhesive~\citep{Cabello2016Soft}. Contrary to the metamaterial inspired interfaces, the bulk adhesive can only offer linear design space in terms of normalised density. While such statement alone might sound insignificant, or even obvious, from a designers perspective this observation is rather striking. Namely, metamaterials allows a very flexible adjustment of size of regions transferring the stresses by either changing the unit cell geometry or density. We also note that, intuitively for the bending dominated case, the hexagonal unit cell should appear as the most compliant, \emph{i.e.} longest characteristic wave-length $\lambda_i$. This is indeed the case, but only over limited range of densities. Once $\bar{\rho}\ge 10^{-1}$, the stretch dominated triangular unit cell becomes the most compliant, with values comparable to soft adhesives ($\square$). Once $\bar{\rho}\ge 3\times 10^{-1}$, the bending dominated interface will become the stiffest, and as stiff as the stiff epoxy adhesive ($\openbigstar$) and similar debonding cases ($\bigcirc$) but at several times lower density.

\begin{figure}
\centering
\subfloat[]{\includegraphics[width=0.5\textwidth]{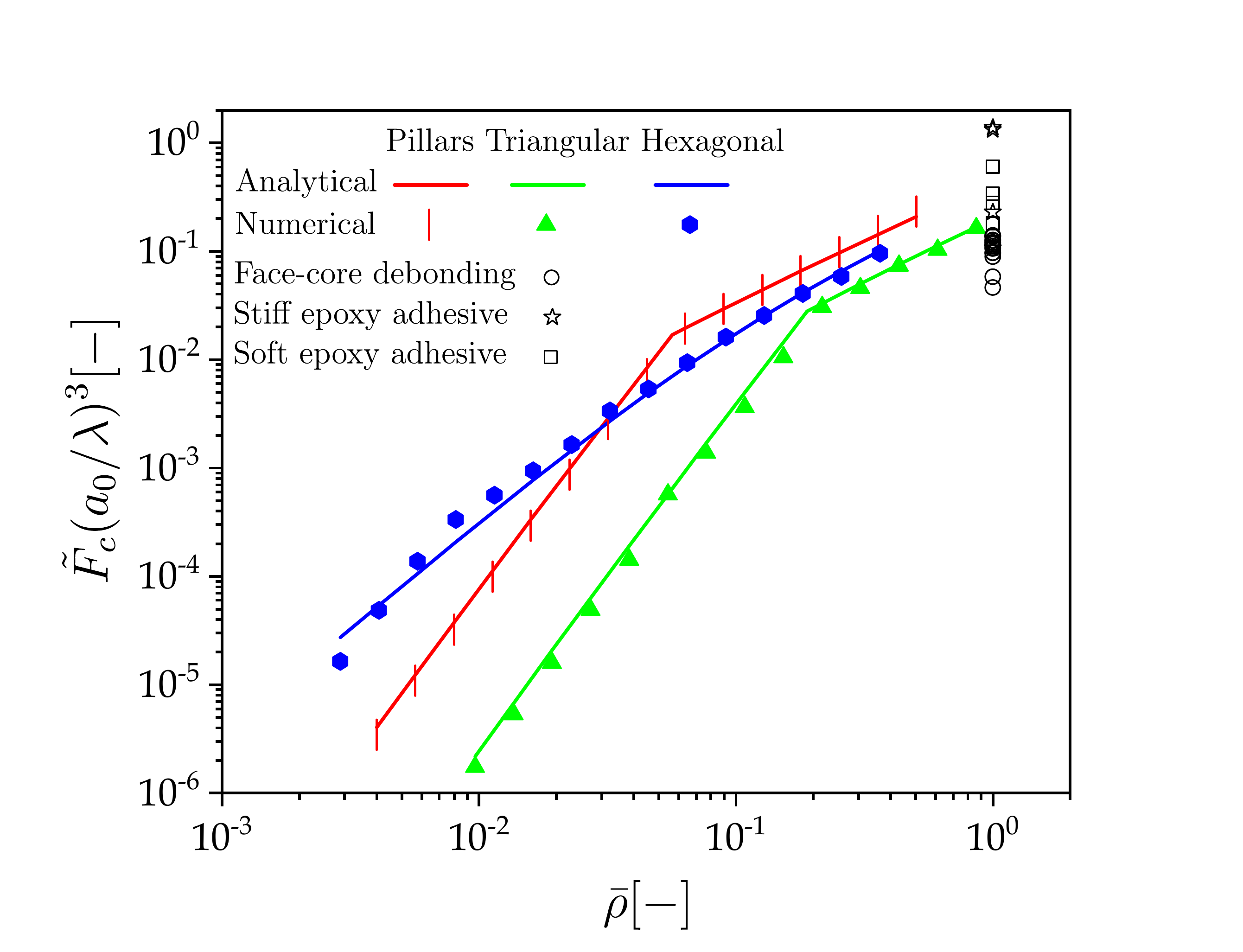}}
\subfloat[]{\includegraphics[width=0.5\textwidth]{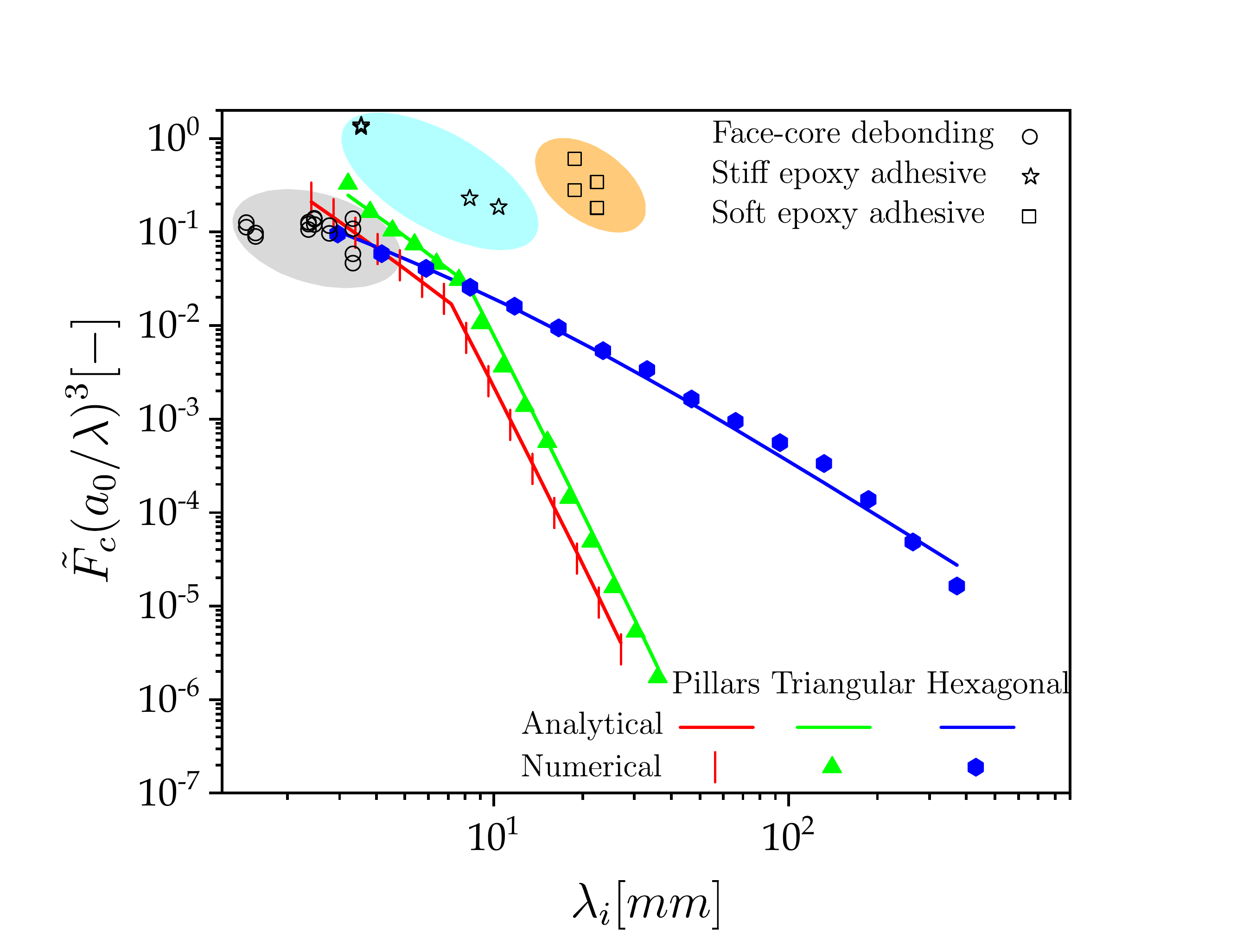}}
\caption{(a) Critical force as a function of the area densities: $\bar\rho_\mathrm{|}=4\tilde{\ell}$, $\bar\rho_\mathrm{\triangle}=4(1+\sqrt{2})\tilde{\ell}$, $\bar\rho_\mathrm{\hexagon}=5\tilde{\ell}/\sqrt{3}$ for $\tilde{h}=1$. (b) Critical force comparison with the wavelength. Here we show the case for $\tilde{h}=1$. Empirical results for sandwich material core debonding ($\bigcirc$), stiff epoxy adhesive ($\openbigstar$) and toughened epoxy adhesive ($\square$) are shown accordingly~\citep{Prasad1994debonding,LopesFernandes2021Stiff,Cabello2016Soft}.}
\label{fig:Master_curves}
\end{figure}

Inspired by the Ashby diagrams~\citep{ashby2017materials}, Fig.~\ref{fig:Master_curves}-(a) gathers the critical forces from the three interfaces here analysed (for clarity, a single case of $\tilde{h}=1$ is presented as representative), and plot them over the range of normalised densities $\bar{\rho}$.  Similarly to what was found in the phase maps, two power-laws are distinguished, depending on the unit cell topology (either stretching or bending dominated) and, within stretching dominated cases, also the failure mode type. However, contrary to previous results, the normalised critical force (per unit length of the process zone) is increasing over the entire span of $\bar{\rho}$.

Like for the previous case, Fig.~\ref{fig:lambda}, the bulk adhesive can only offer vertical design space in terms of normalised density. Moreover, affecting the failure load requires use of different materials, often incompatible with chemical standpoint or with unwanted physical properties (for instance, lower glass transition temperature etc.), and offer only limited span of properties. The metamaterial interfaces have potential to bring new capabilities by using single material, while limiting amount of the material usage. However, we note, that specifically within the compressive failure regimes, the critical failure force drops faster than the relative density.

\com{One of the existing approaches used to mitigate stress concentrations emerging from geometrical and material discontinuities inherent for bondlines is through grading material properties~\citep{dos2020numerical}. Regarding the matter of materials, a number of issues related to physical and chemical compatibility and stability need to be addressed~\citep{marques2021overview}. Eq.~\eqref{eq:winkler3} indicates, in a generic manner, the relationship between the effective material properties and length of the process zone, $\lambda_i$, over which stresses are distributed. The lower modulus of the interface material will result in a longer stress distribution zone---on the other hand, the more rigid the material of adhesive is, the shorter the process zone length. Therefore, $\lambda_i$ can be used to indicate how well a given material system will cope with transferring the loading over bigger regions. The crucial paradigm shift once in using metamaterial interfaces is related to the fact that $\lambda_i$ can be fully controlled by the geometry of the unit cells. Therefore, all important parameters and properties are selected by a designer making the choice over geometries used}. In Fig.~\ref{fig:Master_curves}-(b) a unique, and to the best of our knowledge not yet reported in the literature, critical failure loads are plotted against the characteristic wave-length $\lambda_i$---this is done for the same cases as the previous results. From this graph we can deduce the relation between the failure loads and extension of process zone. \com{This comparison provides evidence of another potentially important aspect of geometrically controlled interfaces: the range of $\lambda_i$'s available stands well beyond of what can be achieved by controlling chemical composition of the bondline materials.}

\section{Conclusion}
In this work, we attempted to bring together the concept of mechanical metamaterials into adhesive bonding technology and answer the question ``can mechanical metamaterials replace adhesives?''. Aiming at answering or at least shed light at this query, we have developed a theoretical and a numerical framework for studying and analysing systems in which mechanical metamaterials are confined between the two joined materials. An amalgamation of competing effects---metamaterial, interface and joined materials---leads to the relevant characteristic length scales in the problem to be intertwined, which is a situation reminiscent to adhesive joints. Contrary to bulk bondlines, for which characteristic lengths scales are uniquely related to the material of choice, the metamaterials approach unveils a spectrum of possibilities behind geometrical manipulations and modern manufacturing technologies. Thus, accounting for failure criterion can be shifted to a design process that encompasses a wider properties-on-demand philosophy---this also leads to the specification of elastic properties, failure modes and types of load to be carried. 

Three 2-dimensional lattice structures, composed of unit cells refereed to as pillar, triangular and hexagonal, have been used to form interface metamaterial-bondline. Despite the fact that we have worked with two-dimensional examples, here, it is expected that concept presented may be naturally generalised to more a realistic three-dimensional physical models. 
In our theoretical approach, we have considered a homogenised model of lattice interfaces---playing the role of an elastic foundation---which is then sandwiched between two beams to form the metamaterial inspired adhesive joint. Such formulation allowed a fast and reliable prediction of failure onset loads with the insights to the micro-structural details of the unit cell. This model has also proven to be very robust and accurate in outlining the importance of the different length scales responsible for the mechanical performance of these new types of adhesive joints. Relationships amongst unit cell geometry, density, process zone length and the failure load, can be readily obtained. To verify results of the theoretical analysis, a numerical model based on custom FEM formulation was devised. In specific, the trusses constituting metamaterial microstructure were modelled as beams with piece-wise strain-stress relation during loading. Such formulation proved very efficient in solving problem at hand and considerably reduced computational costs related to nowadays often seen formulations using two and three-dimensional FEM. A very good agreement, both quantitative and qualitative between the two approaches is recorded. Both models predict remarkably accurate failure loads and stress fields. In addition, both models predicts the same power-law relations between the failure load and aspect ratio of unit cell truss elements for all the configurations investigated. The results obtained indicate a strong relation between the failure loads and the process zone length. The process zone length is then very sensitive to truss element aspect ratio and unit cell geometry. This novel microstructure geometry driven materials offers a new tool to designers for shaping properties of structure with interfaces.

This work indicates that metamaterial interfaces can potentially lead to a new approach for designing of adhesive joints. In particular, different unit cells can be used to mitigate stress fields over different regions of the joint---thus, the size effects could be controlled. Very low weight-to-volume ratios could be achieved with controlled lost of mechanical performance. However, the present work is by no means conclusive, as it is our intention and hopes that this novel approach will provoke fresh discussion on alternatives into design of adhesive joints and interfaces between materials.











\bibliographystyle{cas-model2-names}
\bibliography{references.bib}

\end{document}